\documentclass[preprint]{ptephy_v1}

\preprintnumber{NITEP 46} 

\usepackage{amsmath}
\usepackage{amssymb}
\usepackage{bm}

\def\beq{\begin{equation}}
\def\eeq{\end{equation}}
\def\bea{\begin{eqnarray}}
\def\eea{\end{eqnarray}}
\def\fun#1#2{\lower3.6pt\vbox{\baselineskip0pt\lineskip.9pt
  \ialign{$\mathsurround=0pt#1\hfil##\hfil$\crcr#2\crcr\sim\crcr}}}

\usepackage{color}
\usepackage{ulem}
\renewcommand\sout{\bgroup\color{red} \ULdepth=-.5ex \ULset}

\begin{document}

\title{Correspondence between isoscalar monopole strengths and $\alpha$ inelastic cross sections on $^{24}$Mg }

\author[1,2,3]{Kazuyuki Ogata}
\affil[1]{Research Center for Nuclear Physics, Osaka University,
Ibaraki, Osaka 567-0047, Japan \email{kazuyuki@rcnp.osaka-u.ac.jp}}
\affil[2]{Department of Physics, Osaka City University, Osaka 558-8585, Japan}
\affil[3]{Nambu Yoichiro Institute of Theoretical and Experimental Physics (NITEP), Osaka City University, Osaka 558-8585, Japan}

\author[2,3]{Yohei Chiba}
\author[2,3]{Yukinori Sakuragi}


\begin{abstract}%
The correspondence between the isoscalar monopole (IS0) transition strengths and $\alpha$ inelastic cross sections, the $B({\rm IS0})$-$(\alpha,\alpha')$ correspondence, is investigated for $^{24}$Mg($\alpha,\alpha'$) at 130 and 386~MeV. We adopt a microscopic coupled-channel reaction framework to link structural inputs, diagonal and transition densities, for $^{24}$Mg obtained with antisymmetrized molecular dynamics to the ($\alpha,\alpha'$) cross sections. We aim at clarifying how the $B({\rm IS0})$-$(\alpha,\alpha')$ correspondence is affected by the nuclear distortion, the in-medium modification to the nucleon-nucleon effective interaction in the scattering process, and the coupled-channels effect. It is found that these effects are significant and the explanation of the $B({\rm IS0})$-$(\alpha,\alpha')$ correspondence in the plane wave limit with the long-wavelength approximation, which is often used, makes no sense. Nevertheless, the $B({\rm IS0})$-$(\alpha,\alpha')$ correspondence tends to remain because of a strong constraint on the transition densities between the ground state and the $0^+$ excited states. The correspondence is found to hold at 386~MeV with an error of about 20\%--30\%, while it is seriously stained at 130~MeV mainly by the strong nuclear distortion. It is also found that when a $0^+$ state that has a different structure from a simple $\alpha$ cluster state is considered, the $B({\rm IS0})$-$(\alpha,\alpha')$ correspondence becomes less valid. For a quantitative discussion on the $\alpha$ clustering in $0^+$ excited states of nuclei, a microscopic description of both the structure and reaction parts will be necessary.
\end{abstract}


\maketitle

\section{Introduction}
\label{sec1}

The nuclear clustering structure, characterized by weakly-interacting subunits inside a nucleus, is one of the fundamental aspects of atomic nuclei. It emerges near the threshold energies in the excitation spectra, as predicted by the Ikeda diagram~\cite{Ike68}, and also in the ground states of several nuclei~\cite{Kan12}. So far, many attempts for directly probing the $\alpha$ cluster states have been done by means of the resonant scattering~\cite{Yam14}, $\alpha$ transfer~\cite{Bec78,Ana79,Oel79,Tan81,Fuk16,Fuk19}, and $\alpha$ knockout processes~\cite{Roo77,Nad80,Car84,Wan85,Yos88,Nad89,Nev08,Mab09,Yos16,Lyu18,Yos18,Lyu19,Yos19,Tan21}. Besides, $\alpha$ inelastic scattering has been utilized to investigate an $\alpha$ cluster structure in excited states of nuclei~\cite{Yon97,Yon99,Yon99a,Ito02,Ito03,Uch04,Li07,Ito13,Bag15,Van15,Gup16,Pea16,Ada18}. As discussed in Refs.~\cite{Suz76,Yam08}, the isoscalar monopole (IS0) operator induces a nodal excitation regarding the coordinate between the constituents of a nucleus, by which nuclear cluster states are strongly and selectively populated.

As it is well known, in the plane wave limit with the long-wavelength approximation, the PW-LW limit, the transition matrix of the inelastic scattering of electron, $(e,e')$ scattering, contains the IS0 transition strengths of nuclei; see also \S\ref{sec21} below. This relation has been applied also to the $\alpha$ inelastic scattering, which is governed by the isoscalar transition. In fact, in Ref.~\cite{CK15}, the IS0 transition strengths $B({\rm IS0})$ for $^{24}$Mg calculated with antisymmetrized molecular dynamics (AMD)~\cite{Kan12,Kan95,Kan95a} were shown to have a good correspondence with those extracted from $\alpha$ inelastic scattering data~\cite{Kaw13}. However, according to the aforementioned explanation in the PW-LW limit, the $\alpha$ inelastic cross sections must be proportional to $q^4$, where $q$ is the momentum transfer, and thus steeply drop off at forward angles. On the other hand, experimental $\alpha$-inelastic cross section data to $0^+$ excited states are peaked at zero degree. Therefore, obviously, the correspondence between $B({\rm IS0})$ and the $\alpha$ inelastic cross sections, the $B({\rm IS0})$-$(\alpha,\alpha')$ correspondence, is non-trivial as explained in the PW-LW limit. It will be important to clarify how the correspondence between $B({\rm IS0})$ and $(e,e')$ scattering data changes when $(\alpha,\alpha')$ scattering is considered. Understanding the mechanism of the $B({\rm IS0})$-$(\alpha,\alpha')$ correspondence for $^{24}$Mg reported in Ref.~\cite{CK15} is also crucial for future studies on $B({\rm IS0})$ for various nuclei using experimental $(\alpha,\alpha')$ cross sections.

In this study, we discuss the $(\alpha,\alpha')$ cross sections on $^{24}$Mg to the $0^+$ excited states at 130 and 386~MeV in view of the $B({\rm IS0})$-$(\alpha,\alpha')$ correspondence. To link the structural wave functions of $^{24}$Mg used in Ref.~\cite{CK15} to reaction observables, we employ a microscopic coupled-channel (MCC) framework~\cite{KYO19,KYO19a,KYO19b,KYO20,KYO20a,KYO20b,KY20,KYO21,KYO21a}, in which CC potentials between $\alpha$ and $^{24}$Mg are obtained from diagonal and transition densities of $^{24}$Mg calculated with AMD and the Melbourne nucleon-nucleon (NN) $g$-matrix interactions. This MCC framework can microscopically treat i) the in-medium modification to the NN effective interaction in scattering processes, ii) the nuclear distortion effect, and iii) the coupled-channel effect among the elastic channel and inelastic channels. We investigate the role of each of these effects on the $B({\rm IS0})$-$(\alpha,\alpha')$ correspondence and thereby clarify the difference between the $(e,e')$ and $(\alpha,\alpha')$ processes in their relations to $B({\rm IS0})$. We then discuss the robustness and limitation of the $B({\rm IS0})$-$(\alpha,\alpha')$ correspondence in the present case.

The construction of this paper is as follows. In \S\ref{form}, after recapitulating the explanation for the relation between $B({\rm IS0})$ and the $\alpha$ inelastic cross sections in the PW-LW limit, we briefly introduce the MCC framework for describing $\alpha$ scattering. In \S\ref{res}, we show numerical results for the $\alpha$ elastic and inelastic cross sections on $^{24}$Mg at 130 and 386~MeV and investigate the proportionality of $B({\rm IS0})$ to the inelastic cross sections. Roles of the in-medium modification to the NN effective interaction, the nuclear distortion effect, and the CC effect are discussed. Finally, a summary is given in \S\ref{sum}.

\section{Theoretical framework}
\label{form}

\subsection{Inelastic scattering of $\alpha$ in the plane wave limit with the long-wavelength approximation}
\label{sec21}

First, following an explanation for $(e,e')$ scattering, we recapitulate how one may expect a relevance between $B({\rm IS0})$ and the $\alpha$ inelastic cross sections. In the plane-wave Born approximation (PWBA) limit, the transition matrix to the $i$th $0^+$ state is given with a double folding model by
\begin{equation}
T_i^{\mathrm{PW}}=\int e^{i\boldsymbol{q}\cdot\boldsymbol{R}}t_{\rm NN}\left(
s\right) \rho^{\alpha}\left( r_{\alpha}\right) \rho_{0_{i}^{+}0_{1}^{+}%
}^{\mathrm{A}}\left( r_{\mathrm{A}}\right) d\boldsymbol{r}_{\mathrm{A}%
}d\boldsymbol{r}_{\alpha}d\boldsymbol{R},
\label{tpw}
\end{equation}
where ${\bm q}$ is the momentum transfer, $r_\alpha$ and $r_{\rm A}$ are the coordinates of a nucleon inside $\alpha$ and the target nucleus A, respectively, and ${\bm R}$ is the coordinate vector of the center-of-mass (c.m.) of $\alpha$ regarding that of A. The spin-parity of A in the ground state is assumed to be $0^+$. An NN effective interaction in free space, $t_{\rm NN}$, is employed as a transition interaction in Eq.~(\ref{tpw}). We denote the relative coordinate of a nucleon in $\alpha$ to a nucleon in A by ${\bm s}$. As usually done, we consider only the central part of $t_{\rm NN}$. The exchange term of $t_{\rm NN}$ is dropped for simplicity. $\rho^{\alpha}$ is a nucleon one-body density of $\alpha$ and $\rho_{0_{i}^{+}0_{1}^{+}}^{\mathrm{A}}$ is a nucleon one-body transition density of $^{24}$Mg from the ground state to the $i$th $0^+$ state. Although it is not shown explicitly, because $\alpha$ contains the same numbers of proton and neutron, only the isoscalar transition is included in Eq.~(\ref{tpw}).

One can rewrite $T_i^{\mathrm{PW}}$ by using
$\boldsymbol{R}=\boldsymbol{r}_{\mathrm{A}}+\boldsymbol{s}-\boldsymbol{r}_{\alpha}$
as
\begin{equation}
T^{\mathrm{PW}}=\tilde{t}_{\rm NN}\left( q\right) \tilde{\rho}^{\alpha}\left(
q\right) \tilde{\rho}_{0_{i}^{+}0_{1}^{+}}^{\mathrm{A}}\left( q\right)
\end{equation}%
with
\begin{equation}
\tilde{t}_{\rm NN}\left( q\right) =
4\pi\int j_{0}\left( qs \right) t_{\rm NN}\left( s\right) s^2 ds,
\end{equation}%
\begin{equation}
\tilde{\rho}^{\alpha}\left( q\right) =
4\pi\int j_{0}\left( qr_{\alpha}\right)
\rho^{\alpha}\left( r_{\alpha}\right) r_{\alpha}^{2}dr_{\alpha},
\end{equation}%
\begin{equation}
\tilde{\rho}_{0_{i}^{+}0_{1}^{+}}^{\mathrm{A}}\left( q\right) =
4\pi\int j_{0}\left( qr_{\mathrm{A}}\right) \rho_{0_{i}^{+}0_{1}^{+}%
}^{\mathrm{A}}\left( r_{\mathrm{A}}\right) r_{\mathrm{A}}^{2}dr_{\mathrm{A}}.
\label{rhoatil}%
\end{equation}
If we make a long-wavelength approximation to the spherical Bessel function $j_0$, we have
\begin{equation}
T^{\mathrm{PW}} \approx
-\frac{8\pi}{3} J_{\rm NN} q^{2}
\int r_{\mathrm{A}}^{2}\rho_{0_{i}^{+}0_{1}^{+}}^{\mathrm{A}}\left(r_{\mathrm{A}}\right) r_{\mathrm{A}}^{2}dr_{\mathrm{A}}
\equiv -\frac{8\pi}{3} J_{\rm NN} q^{2} M_i ({\rm IS0})
\end{equation}for $i \neq 1$, where $J_{\rm NN}$ is the volume integral of $t_{\rm NN}$. Note that the 0th order term of $j_{0}$ in Eq.~(\ref{rhoatil}) has no contribution to the inelastic scattering because of the orthogonality of the wave function of $^{24}$Mg. Consequently, the $\alpha$ inelastic cross section reads
\begin{equation}
\frac{d\sigma_i}{d\Omega} \propto
q^4 |M_i ({\rm IS0})|^2 \equiv q^4 B_i({\rm IS0}),
\label{cspwba}
\end{equation}
where $B_i({\rm IS0})$ is the IS0 transition strength to the $0_i^+$ state. Equation~(\ref{cspwba}) has been used in many places to guarantee that $B_i({\rm IS0})$ can be extracted from $\alpha$ inelastic cross sections to $0^+$ excited states.

According to Eq.~(\ref{cspwba}),
one may expect ${d\sigma_i}/{d\Omega}\sim 0$ at very forward angles.
To be precise, because of the excitation energy $\epsilon$ of $^{24}$Mg,
$q$ remains finite even at $\theta=0$.
At any rate, however, the inelastic cross sections to $0^+$ excited states of A should steeply drop off when $\theta \to 0$. This is indeed the case with ($e,e'$) scattering. On the other hand, $(\alpha,\alpha')$ cross sections to $0^+$ excited states are even peaked at $\theta=0$ (see, e.g., Fig.~\ref{fig1}(b) and \ref{fig2}(b) below).

\subsection{Microscopic coupled-channel framework for describing $\alpha$-$^{24}$Mg scattering}
\label{sec22}

In the present study, we adopt an MCC framework for $\alpha$ inelastic scattering as in Refs.~\cite{KYO19,KYO19a,KYO19b,KYO20,KYO20a,KYO20b,KY20,KYO21,KYO21a}. We prepare one-body diagonal and transition densities of $^{24}$Mg by using AMD with the Gogny D1S interaction~\cite{Ber91} and construct CC potentials of $\alpha$ with the extended nucleon-nucleus folding (NAF) model. The AMD results for the IS0 transition of $^{24}$Mg for $\epsilon=9$--30~MeV were discussed in detail in Ref.~\cite{CK15}. In this study. we concentrate on the $0^+$ states below the giant monopole resonance region ($\epsilon < 15$~MeV) having $B({\rm IS0})$ larger than about 10\% of the value for the $0_2^+$ state. On top of that, to discuss a possible CC effect, we take into account three low-lying $2^+$ states. Thus, we include the $0_{1,2,3,5,7,8}^+$ and $2_{1,2,3}^+$ states in the CC calculation. As for the energies of these levels, we employ experimental values.

The NAF model was proposed to describe $\alpha$ elastic scattering in Ref.~\cite{Ega14} and its extended version to inelastic scattering, the extended NAF model, has successfully been applied to proton and $\alpha$ scattering on several nuclei~\cite{KYO19,KYO19a,KYO19b,KYO20,KYO20a,KYO20b,KY20,KYO21,KYO21a}. In the extended NAF model, first, diagonal and transition potentials for a nucleon-nucleus (NA) scattering are calculated with a single-folding model; the localization method proposed by Brieva and Rook~\cite{BR77} is applied to the exchange part. Then the $\alpha$-nucleus potentials are obtained by folding the NA potentials with a one-body density of $\alpha$. Justification and advantages of this approach can be found in Refs.~\cite{Ega14,KYO19}.

Following Refs.~\cite{KYO19,KYO19a,KYO19b,KYO20,KYO20a,KYO20b,KY20,KYO21,KYO21a}, we multiply the transition densities calculated with AMD by a factor to reproduce experimental values of $B({\rm IS0})$ and $B({E2})$. The $^4$He one-body density of the one-range Gaussian form given by Eq.~(24) of Ref.~\cite{SL79} is adopted.
As for the effective NN interaction for the extended NAF model, we employ the Melbourne $g$-matrix interaction~\cite{Amo00} based on the Bonn-B potential~\cite{Mac87} evaluated in infinite nuclear matter. The density dependence of the $g$ matrix is taken into account with the local density approximation.
When we discuss the role of the in-medium modification to the NN effective interaction, we use the NN $t$-matrix interaction by Franey and Love (FL)~\cite{FL85} instead. To see the distortion effect, a PWBA calculation is performed by neglecting all the $\alpha$-nucleus scattering potentials; the transition interaction in the PWBA transition matrix is assumed to be the Melbourne $g$ matrix or FL $t$ matrix. Similarly, the role of the CC effect is investigated by making a DWBA calculation; the couplings among the inelastic channels and the back-couplings to the elastic channel are disregarded.

\section{Results and discussion}
\label{res}

\subsection{Elastic and inelastic scattering cross sections of $\alpha$ on $^{24}$Mg}
\label{sec31}
First, we demonstrate how the present MCC framework describes the experimental data for the $\alpha$-$^{24}$Mg elastic scattering and inelastic scattering to the $0_2^+$ and $2_1^+$ states at 130 and 386~MeV. Even though the main purpose of this study is not to reproduce the data, it will be preferable to adopt a microscopic framework that is reasonably consistent with experimental results.

\begin{table}[hbtp]
 \caption{$B(E0)$ and $B(E2)$ values in the unit of $e^2 {\rm fm}^4$ obtained by AMD compared with experimental data~\cite{End79,End93}. Renormalization factors $f_{\rm tr}$ for the transition densities are also shown.}
 \label{tab1}
 \centering
 \begin{tabular}{ccccc}
        & $J^\pi$ &  AMD   & expt.        & $f_{\rm tr}$ \\
  \hline
  $E0$  & $0_2^+$ &  30.32 & $44.9 \pm 5.4$ & 1.217        \\
  $E2$  & $2_1^+$ & 467.24 & $426  \pm 9  $ & 0.955        \\
  $E2$  & $2_2^+$ &  34.21 & $ 33  \pm 2  $ & 0.955        \\
  $E2$  & $2_3^+$ &  24.59 & $11.6 \pm 4.6$ & 0.687        \\
 \end{tabular}
\end{table}
%
In Table~\ref{tab1}, we summarize the $B({E0})$ and $B({E2})$ values for the states included in this study. Available experimental data are shown as well as the renormalization factor for the transition density. Except for this renormalization, we do not include any adjustable parameters as in the preceding studies ~\cite{KYO19,KYO19a,KYO19b,KYO20,KYO20a,KYO20b,KY20,KYO21,KYO21a}.

\begin{figure}[htpb]
\begin{center}
\includegraphics[width=0.48\textwidth,clip]{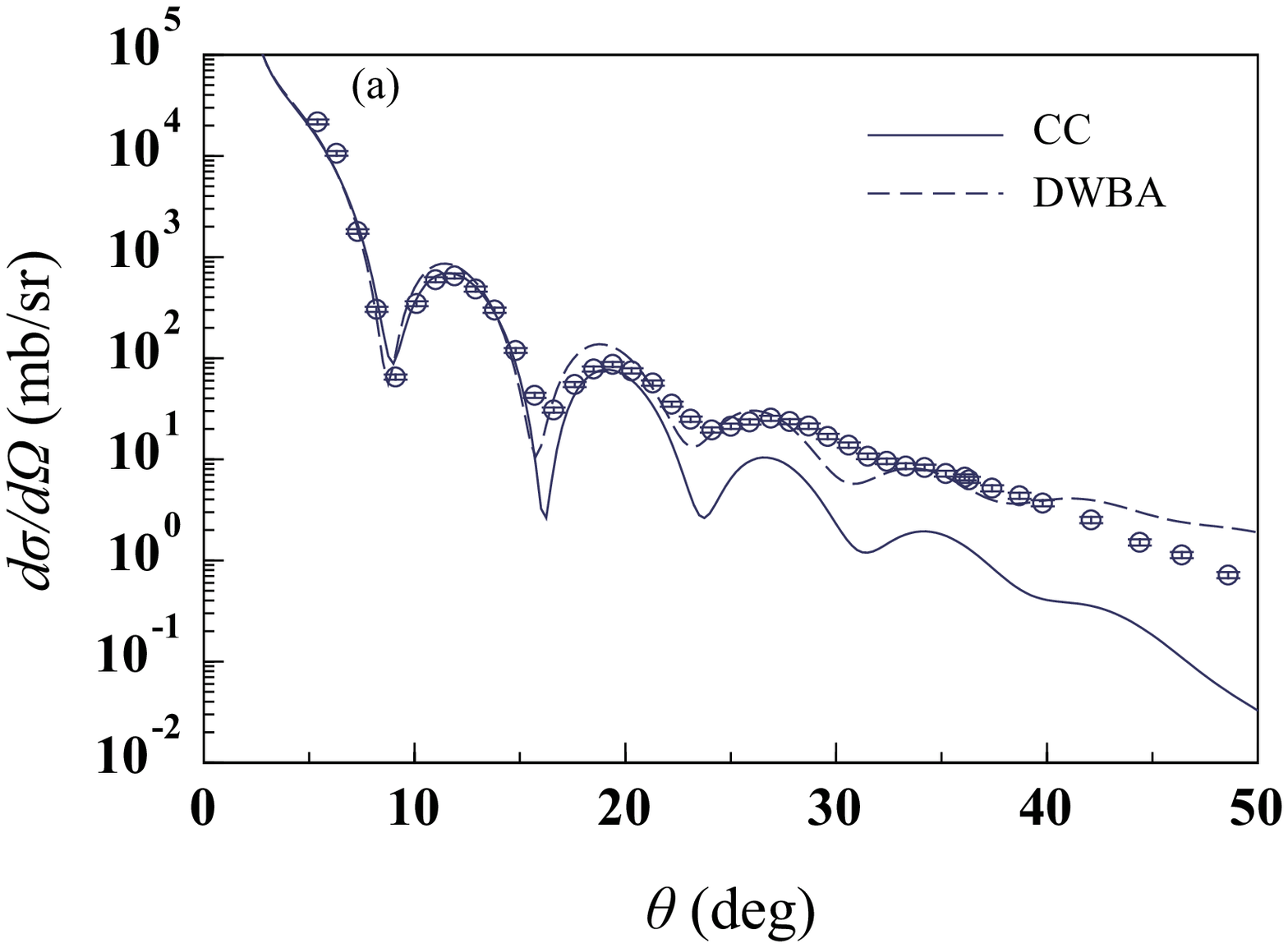}
\includegraphics[width=0.48\textwidth,clip]{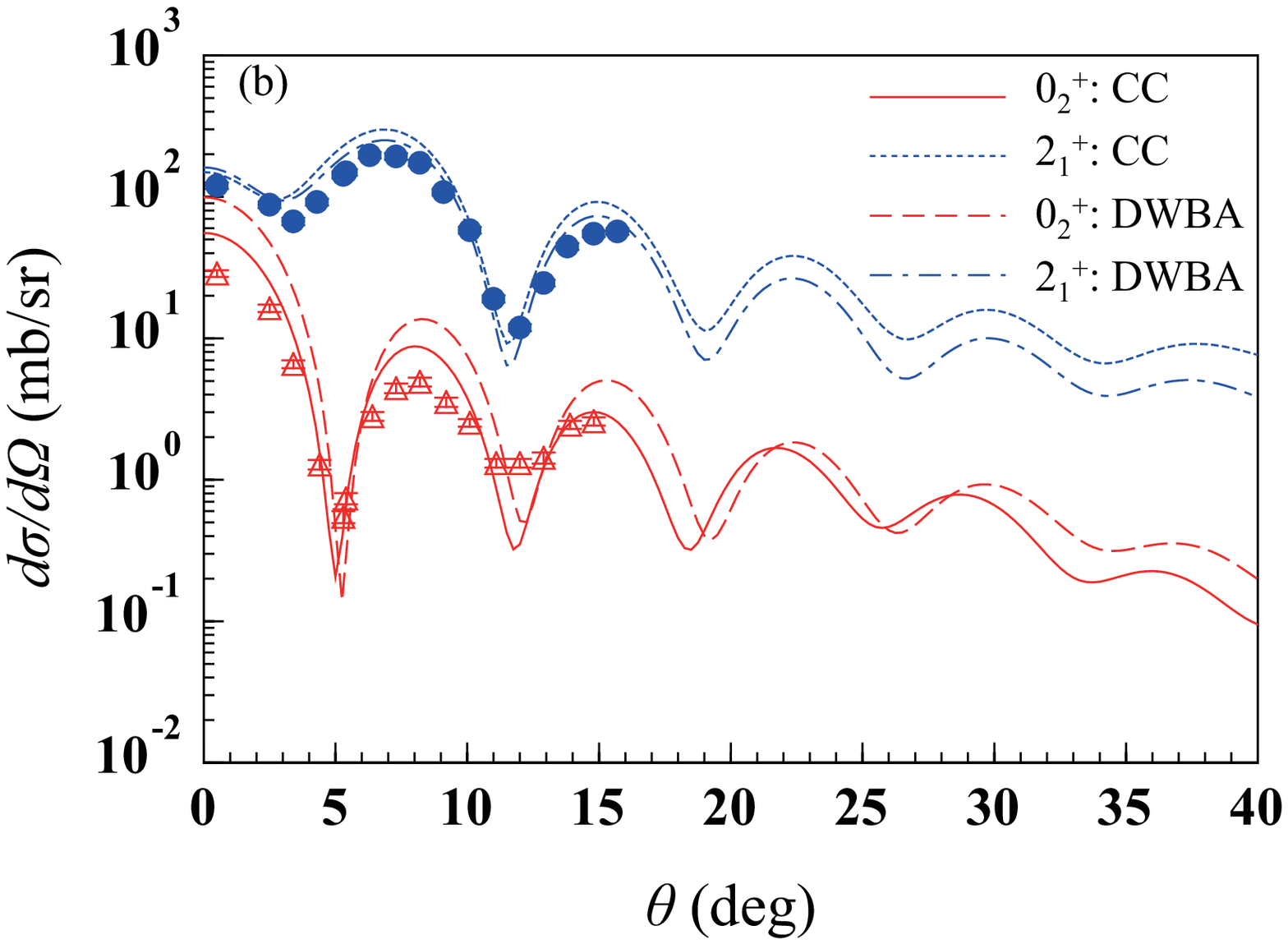}
\caption{(a) $\alpha$ elastic cross section on $^{24}$Mg at 130~MeV with respect to the c.m. scattering angle $\theta$. The solid and dashed lines represent the results of the CC and DWBA calculations, respectively. (b) Angular distribution of the $\alpha$ inelastic cross section on $^{24}$Mg at 386~MeV. The solid (dashed) and dotted (dot-dashed) lines show the results of the CC (DWBA) calculation to the $0_2^+$ and $2_1^+$ states, respectively. Experimental data are taken from Ref.~\cite{Ada18}.}
\label{fig1}
\end{center}
\end{figure}
%

\begin{figure}[htpb]
\begin{center}
\includegraphics[width=0.48\textwidth,clip]{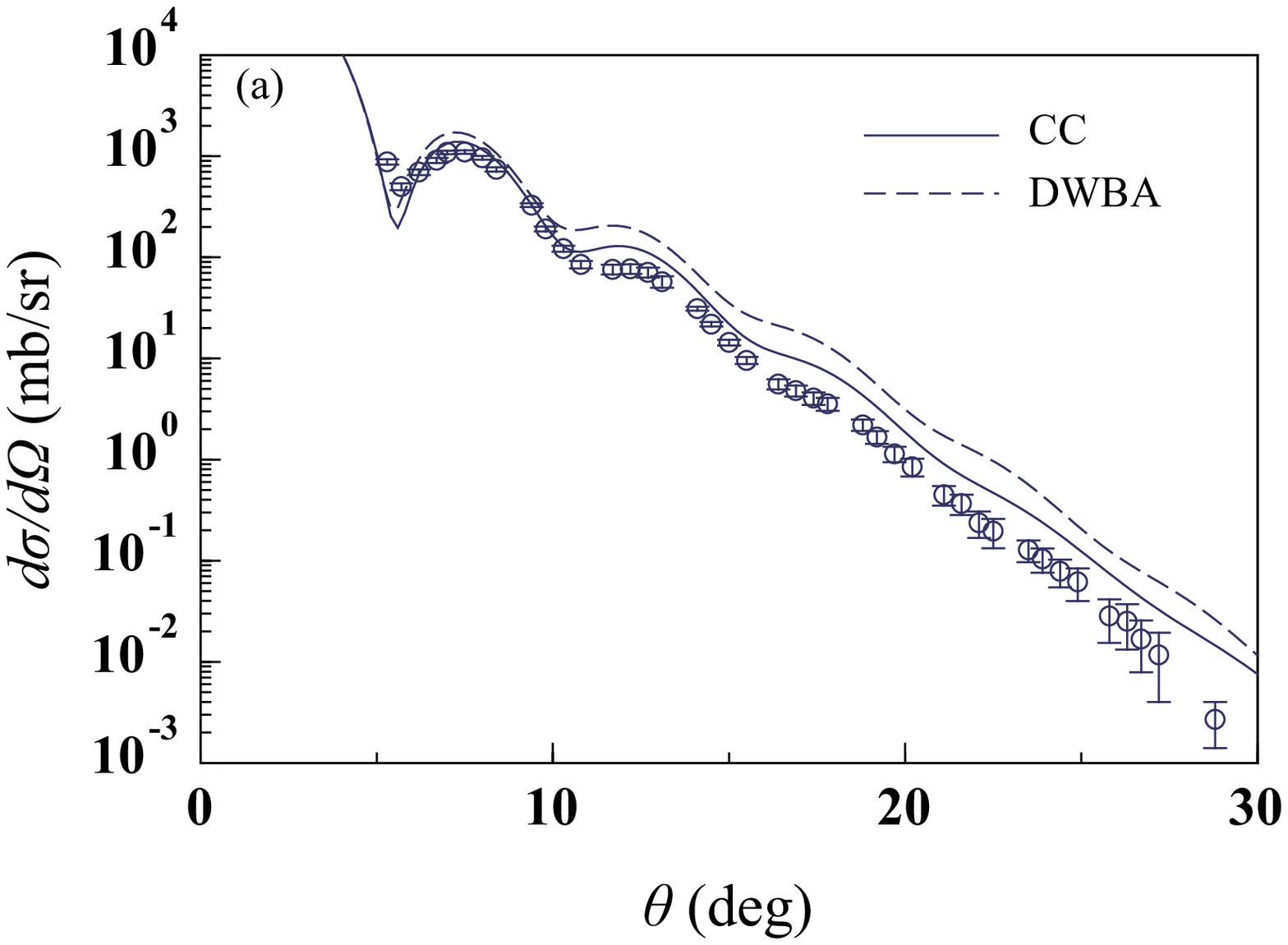}
\includegraphics[width=0.48\textwidth,clip]{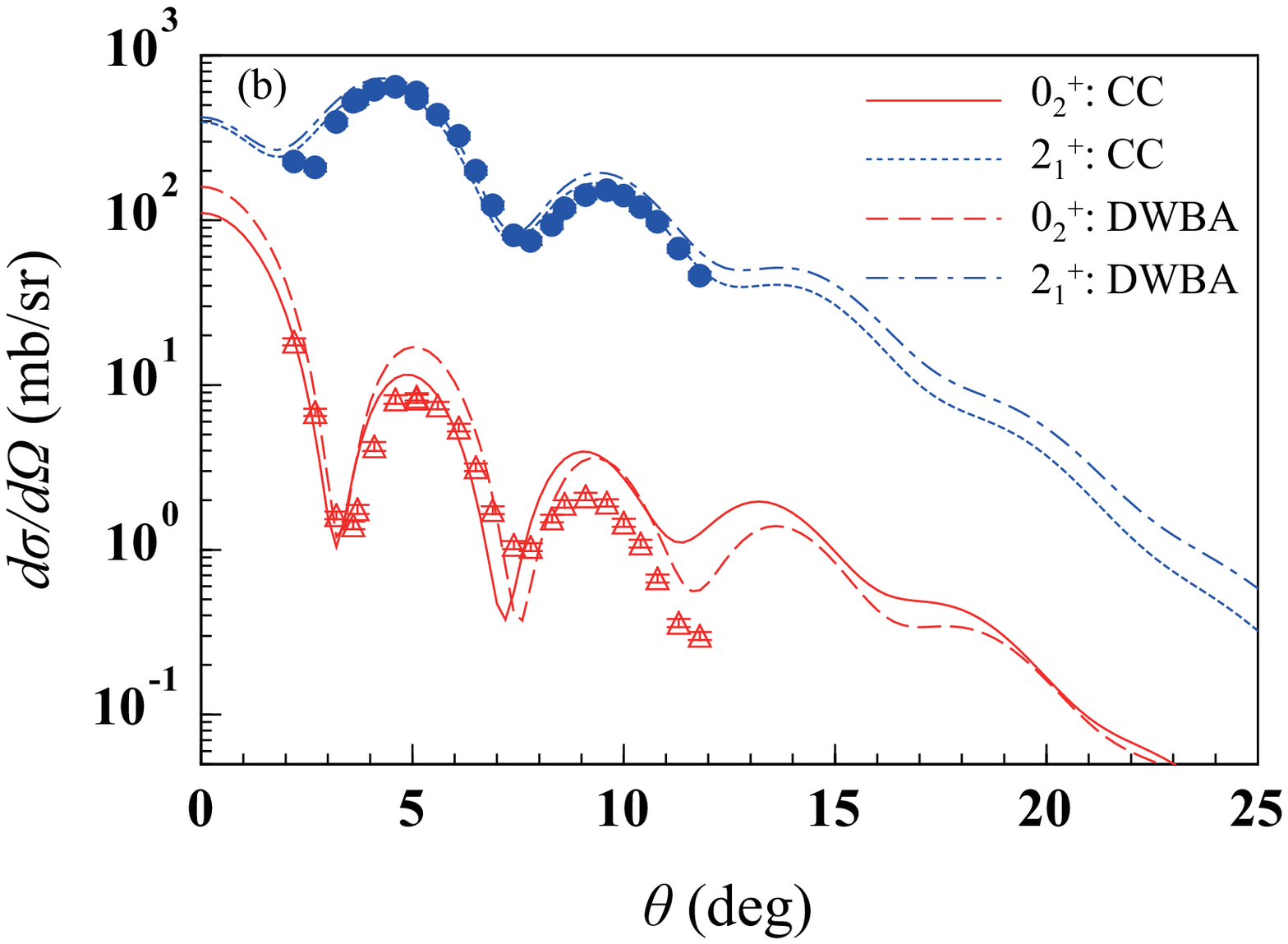}
\caption{Same as Fig.~\ref{fig1} but at 386~MeV.}
\label{fig2}
\end{center}
\end{figure}
We show in Fig.~\ref{fig1}(a) the $\alpha$ elastic scattering cross section at 130~MeV as a function of the c.m. scattering angle $\theta$. The solid (dashed) line shows the result of the CC (DWBA) calculation. Figure~\ref{fig1}(b) shows the inelastic cross sections to the $0_2^+$ and $2_1^+$ states. The solid ($0_2^+$) and dotted ($2_1^+$) lines correspond to the CC calculation, whereas the dashed ($0_2^+$) and dot-dashed ($2_1^+$) lines to the DWBA calculation. Experimental data are taken from Ref.~\cite{Ada18}. The results at 386~MeV are shown in Fig.~\ref{fig2}.

From Figs.~\ref{fig1} and \ref{fig2}, we conclude that the current MCC calculation reproduces reasonably well the experimental data. Although some deviation remains in the elastic cross section at backward angles and in the $0^+_2$ cross section around some peaks, we have not included further adjustable parameters in the reaction part. The results in Figs.~\ref{fig1} and \ref{fig2} will guarantee that a sufficiently meaningful discussion on the $B({\rm IS0})$-$(\alpha,\alpha')$ correspondence can be done with the structure and reaction models adopted in the present study. A systematic study on the microscopic description of $\alpha$ inelastic scattering on $^{24}$Mg at several energies is reported in Refs.~\cite{KYO21,KYO21a}.

One sees from Figs.~\ref{fig1}(b) and \ref{fig2}(b) that while the CC effect on the $2_1^+$ cross sections is minor, it is sizable for the $0_2^+$ cross sections at not only 130~MeV but also 386~MeV. This effect should not be neglected for a quantitative discussion on the $\alpha$ cluster structure in the $0_2^+$ state of $^{24}$Mg. We will return to this point in \S\ref{sec34}.

\subsection{Angular distribution of the $\alpha$-$^{24}$Mg inelastic cross section}
\label{sec32}

%
\begin{figure}[htpb]
\begin{center}
\includegraphics[width=0.48\textwidth,clip]{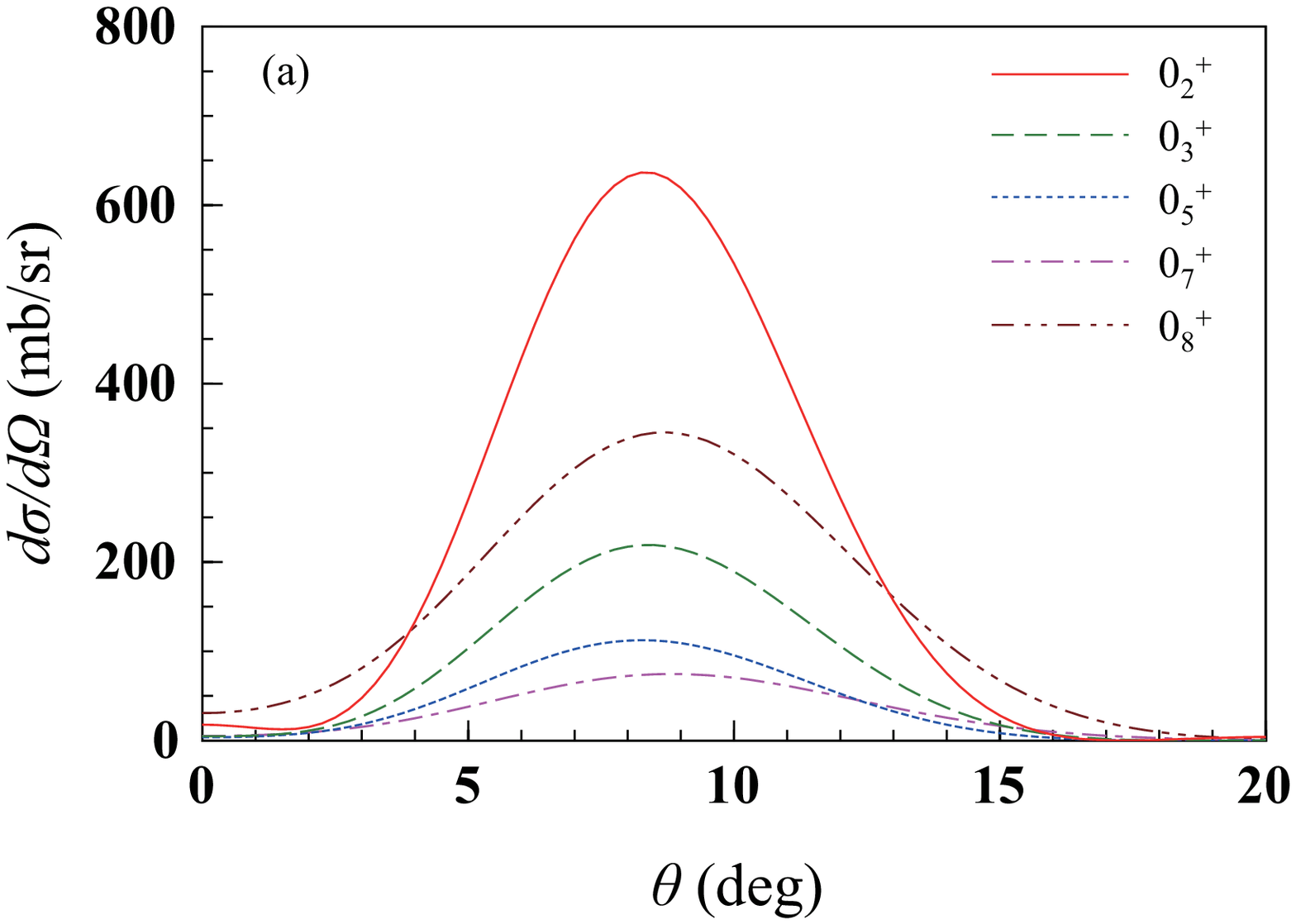}
\includegraphics[width=0.48\textwidth,clip]{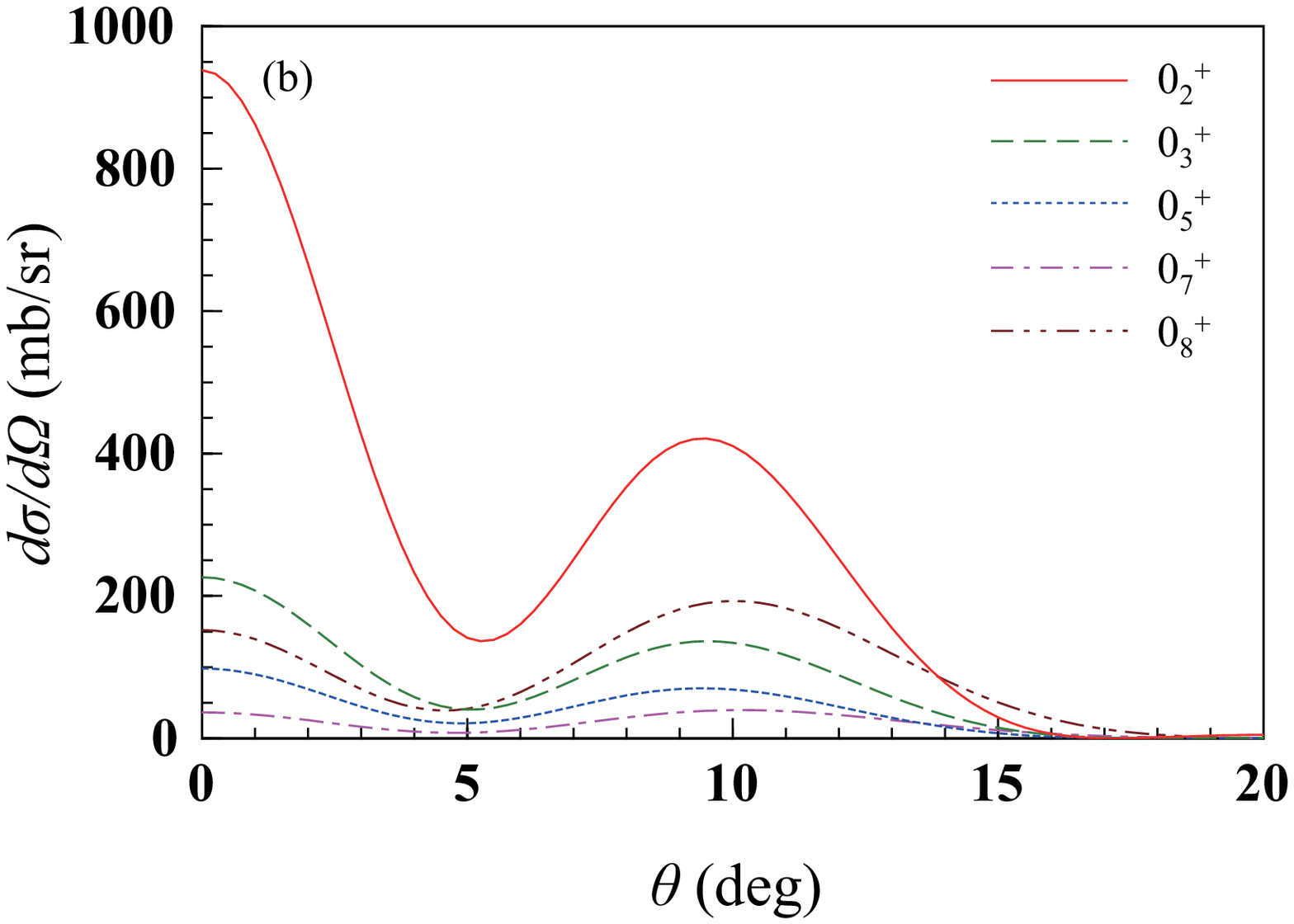}
\includegraphics[width=0.48\textwidth,clip]{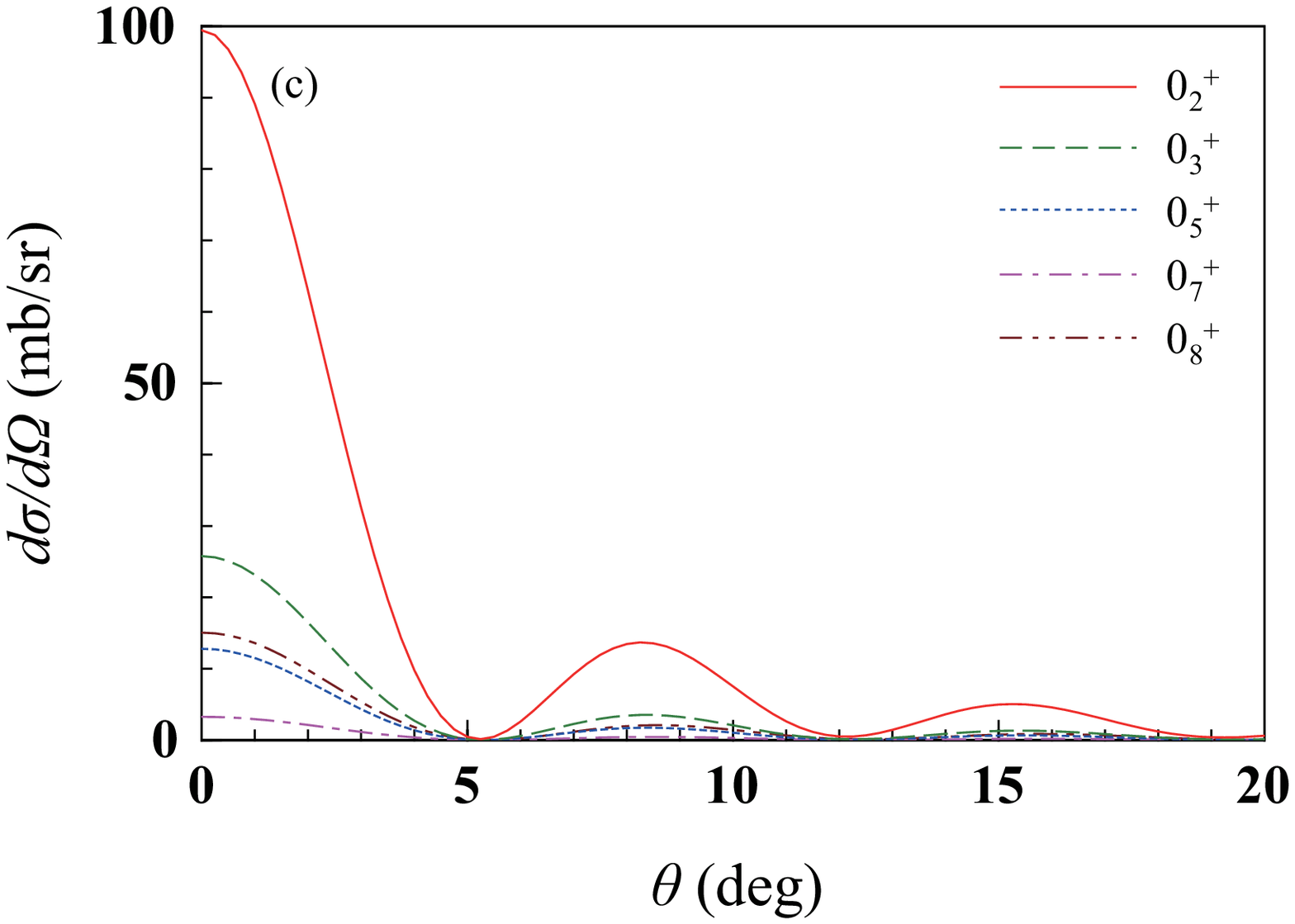}
\includegraphics[width=0.48\textwidth,clip]{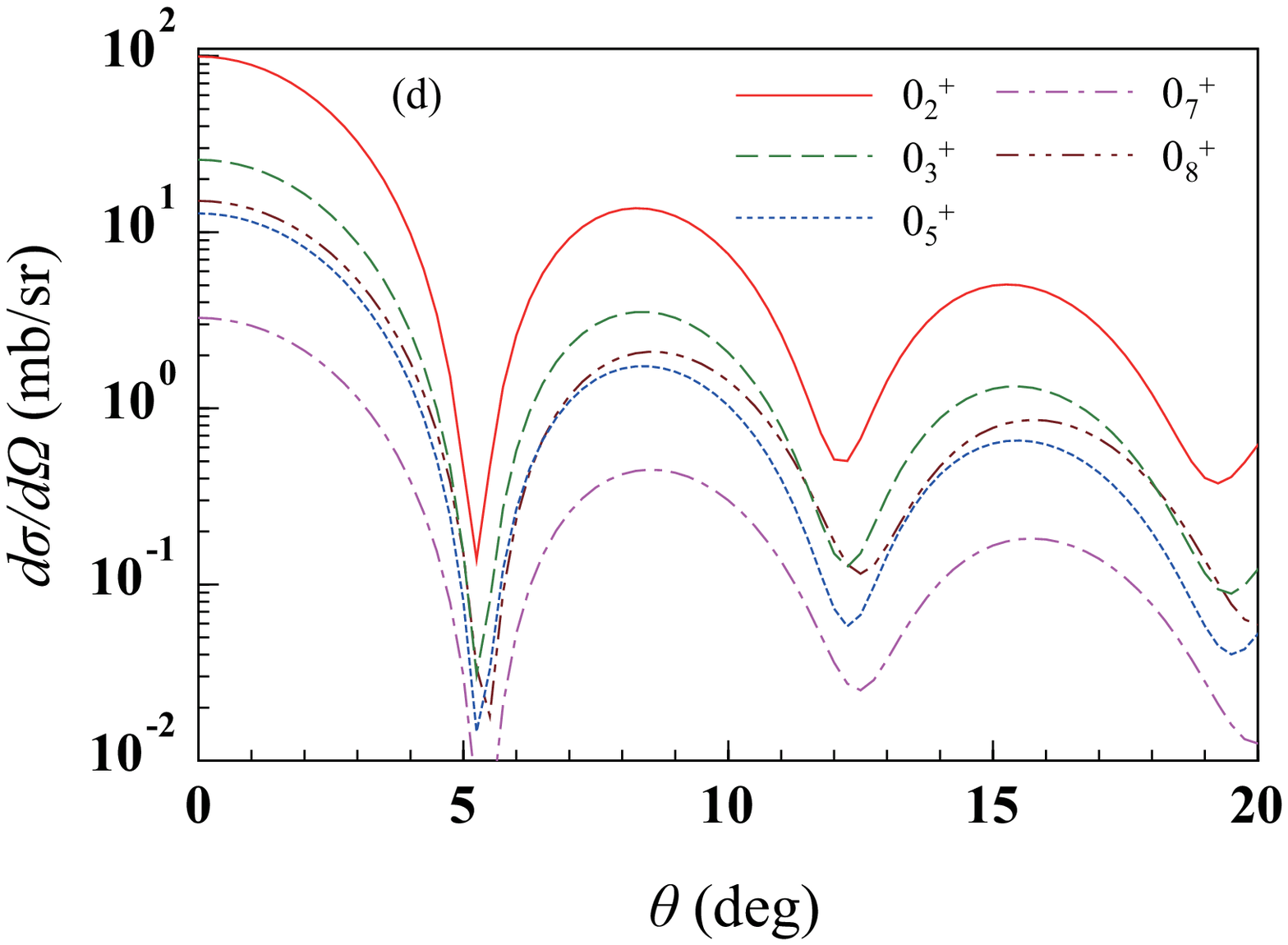}
\caption{$\alpha$ inelastic cross section on $^{24}$Mg at 130~MeV obtained with (a) PWBA-$t$, (b) PWBA-$g$, and (c) DWBA-$g$; (d) is the same as (c) but on the logarithmic scale. In each panel, the solid, dashed, dotted, dot-dashed, dot-dot-dashed lines correspond to the cross sections to the $0_2^+$, $0_3^+$, $0_5^+$, $0_7^+$, and $0_8^+$ states, respectively.}
\label{fig3}
\end{center}
\end{figure}
\begin{figure}[htpb]
\begin{center}
\includegraphics[width=0.48\textwidth,clip]{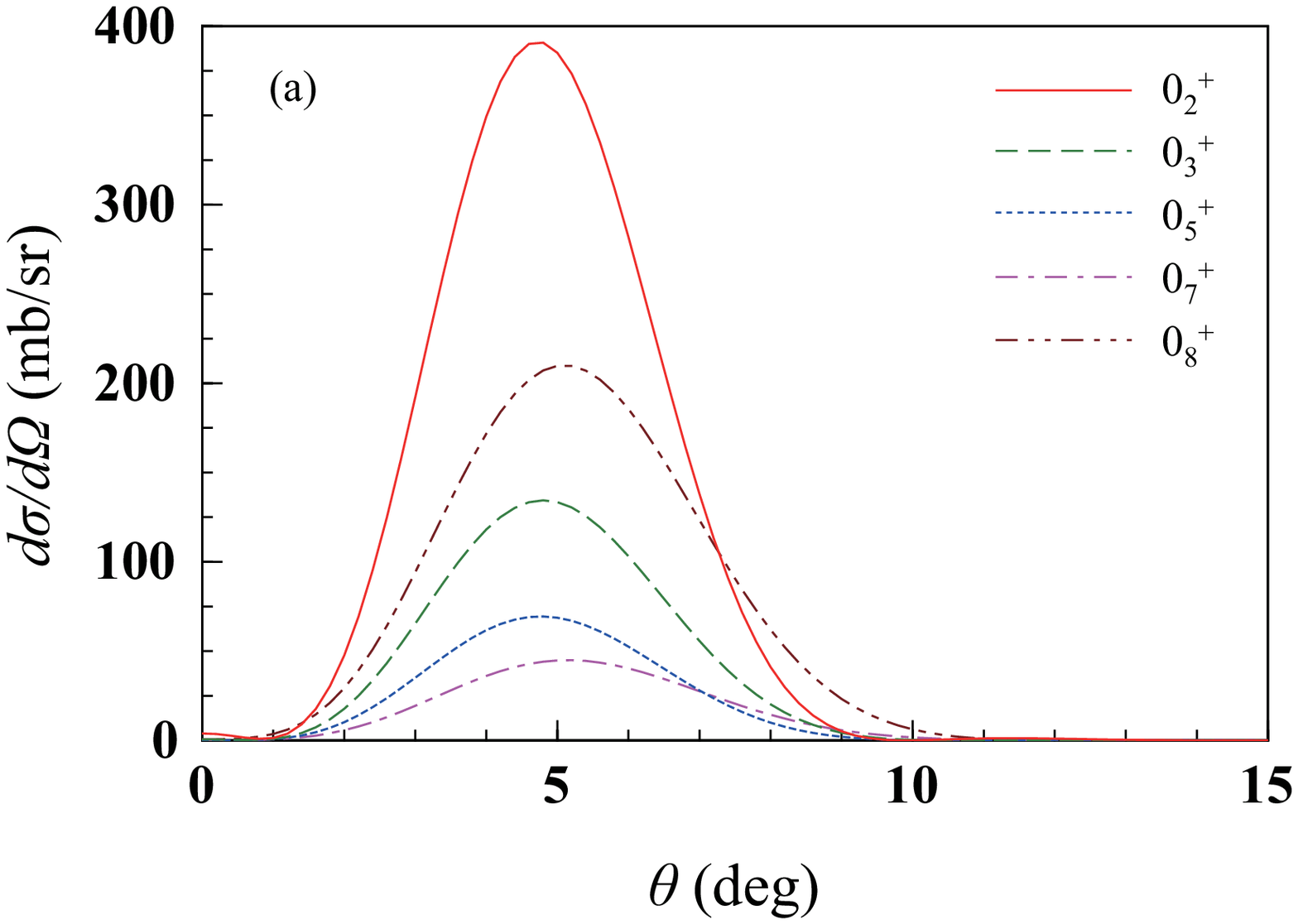}
\includegraphics[width=0.48\textwidth,clip]{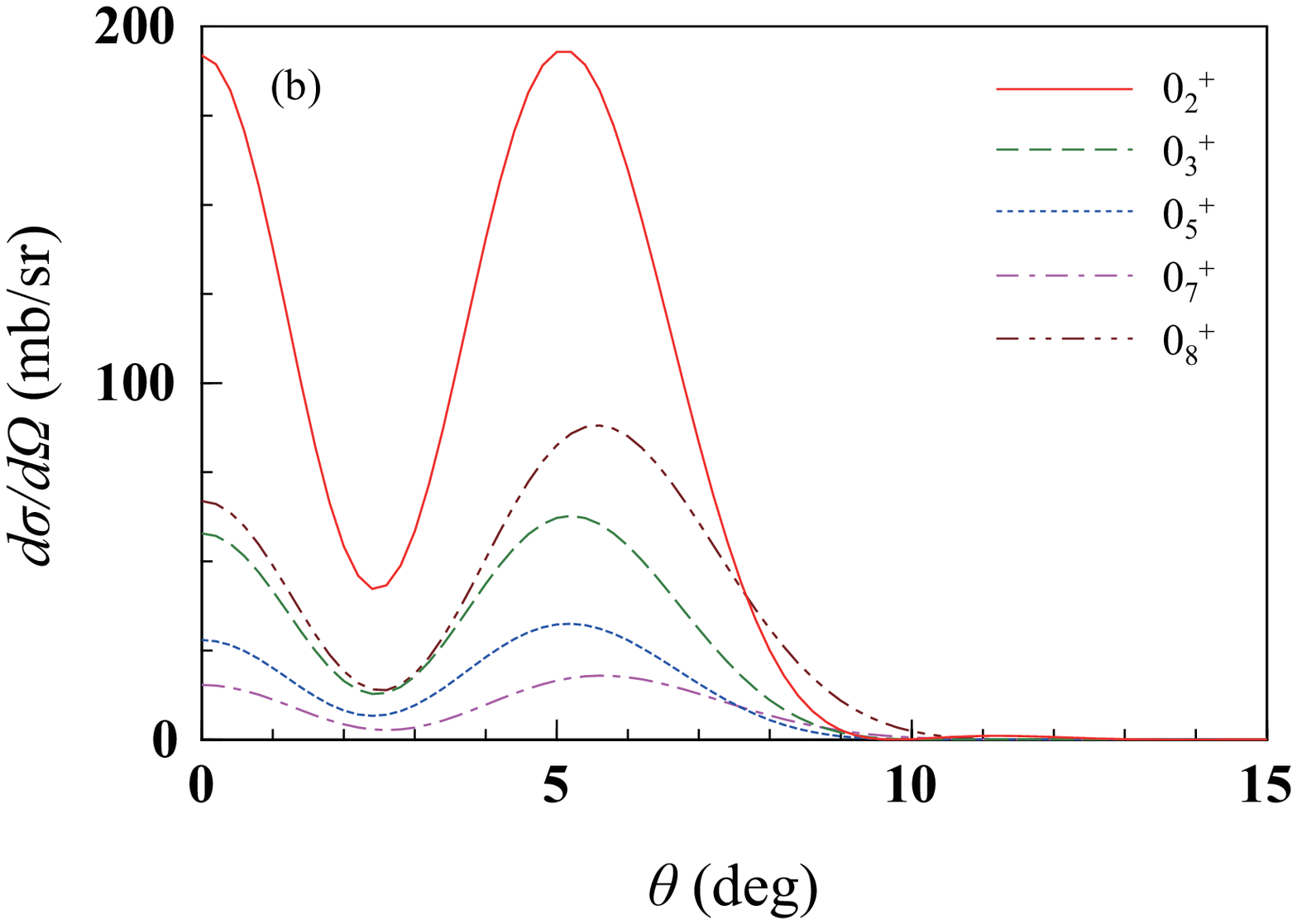}
\includegraphics[width=0.48\textwidth,clip]{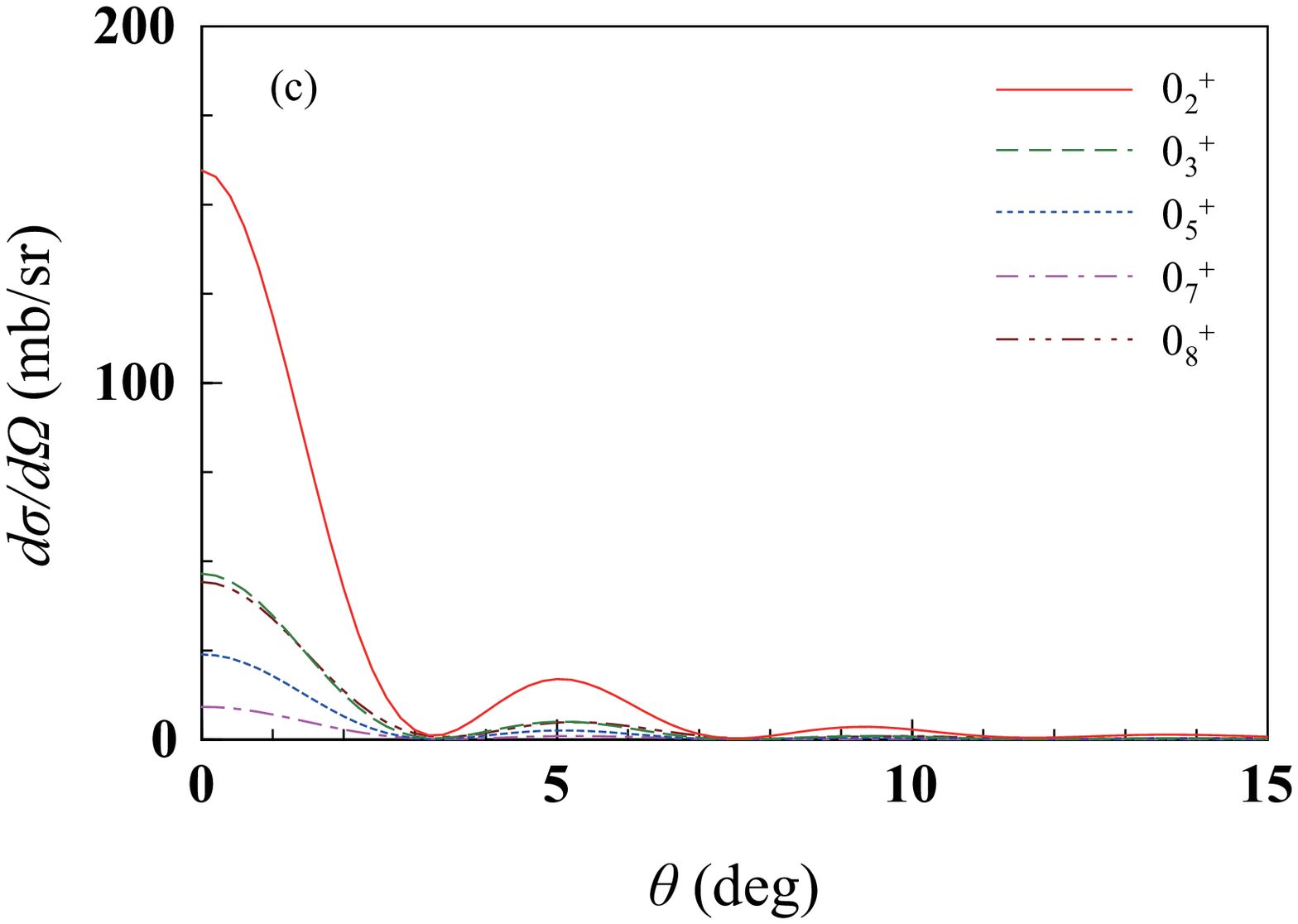}
\includegraphics[width=0.48\textwidth,clip]{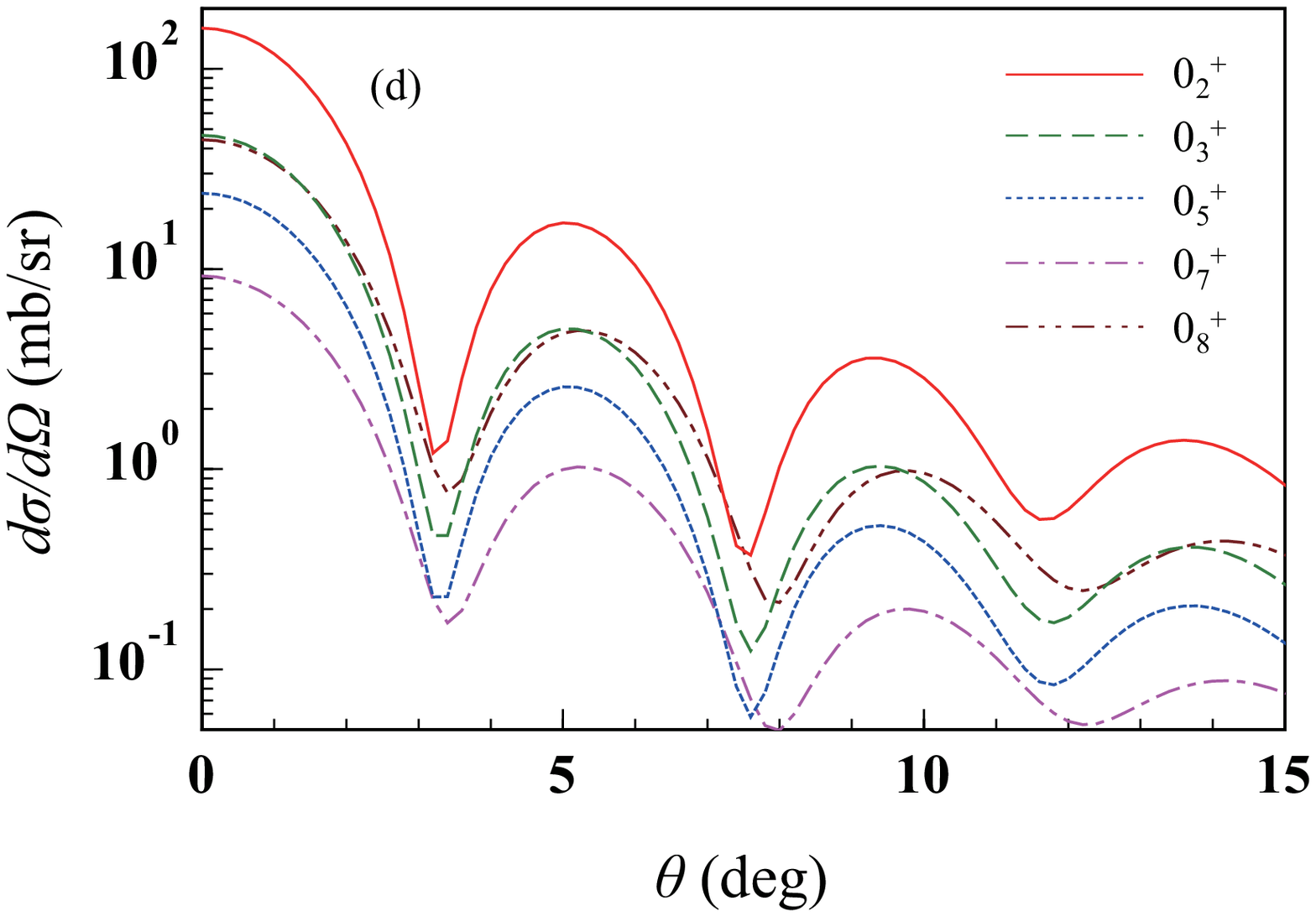}
\caption{Same as Fig.~\ref{fig3} but at 386~MeV.}
\label{fig4}
\end{center}
\end{figure}

Next, we discuss the relation between $B({\rm IS0})_i$ and ${d\sigma_i}/{d\Omega}$ for $i=2$, 3, 5, 7, and 8 within the PWBA and DWBA frameworks. The CC effect on it is discussed in \S\ref{sec34}. In Figs.~\ref{fig3}(a), ~\ref{fig3}(b), and ~\ref{fig3}(c), respectively, we show ${d\sigma_i}/{d\Omega}$ at 386~MeV calculated with PWBA using the FL $t$ matrix (PWBA-$t$), PWBA with the Melbourne $g$ matrix (PWBA-$g$), and DWBA with the Melbourne $g$ matrix (DWBA-$g$). Figure~\ref{fig3}(d) is the same as Fig.~\ref{fig3}(c) but on the logarithmic scale. In each panel, the solid, dashed, dotted, dot-dashed, and dot-dot-dashed lines correspond to $i=2$, 3, 5, 7, and 8, respectively. One sees a clear difference in the shape of the cross section between Figs.~\ref{fig3}(a), \ref{fig3}(b), and \ref{fig3}(c). With PWBA-$t$, the cross section has a peak around $8^\circ$ and decreases as $\theta$ tends to 0. This is the behavior that we discussed in \S\ref{sec21}. In other words, if the $\alpha$ inelastic cross section had the shape shown in Fig.~\ref{fig3}(a), the $B({\rm IS0})$-$(\alpha,\alpha')$ correspondence would be the same as that for $(e,e')$ scattering. However, when we include the in-medium modification to the NN effective interaction (PWBA-$g$), as shown in Fig.~\ref{fig3}(b), a peak at $\theta=0$ newly appears for each $i$. This indicates how strongly the $\alpha$ inelastic cross section at $\theta=0^\circ$ is sensitive to the detail of the transition interaction. Then, by including the nuclear distortion, as seen from Figs.~\ref{fig3}(c) and \ref{fig3}(d), the peaks at $0^\circ$ are more emphasized; one may notice that the absorption effect on the cross sections at $0^\circ$ is considerably weaker than at the second peak around $8^\circ$. Thus, one can understand how the typical shape of the $\alpha$ inelastic cross sections to $0^+$ excited states is developed. Obviously, the $B({\rm IS0})$-$(\alpha,\alpha')$ correspondence cannot be explained in a manner mentioned in \S\ref{sec21}. The results at 386~MeV shown in Fig.~\ref{fig4} have the same features as at 130~MeV.
%
%

%
\begin{figure}[htpb]
\begin{center}
\includegraphics[width=0.6\textwidth,clip]{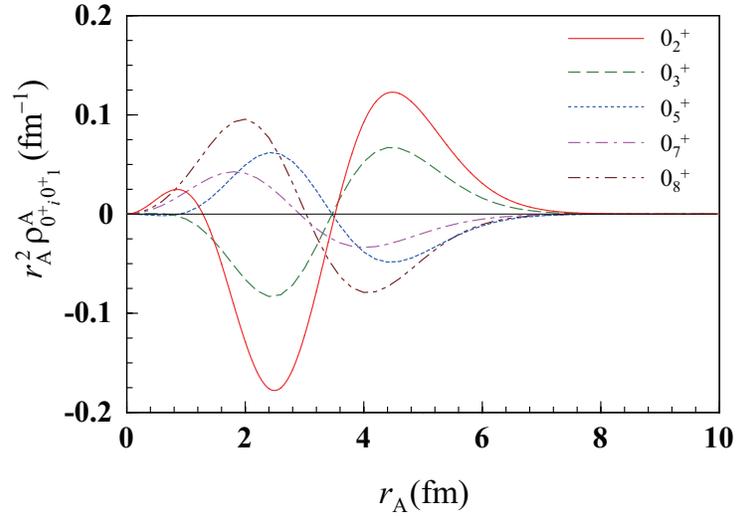}
\caption{Transition densities multiplied by $r_{\rm A}^2$ between the $0_1^+$ and $0_i^+$ states for $i=2$ (solid), 3 (dashed), 5 (dotted), 7 (dot-dashed), and 8 (dot-dot-dashed).}
\label{fig5}
\end{center}
\end{figure}
Below we try to intuitively understand the mechanism of this rather drastic change in the $\alpha$ inelastic cross sections by using the property of the monopole transition density $\rho_{0_{i}^{+}0_{1}^{+}}^{\mathrm{A}}$ shown in Fig.~\ref{fig5}. The meaning of the lines is the same as in Fig.~\ref{fig3}. A characteristic feature of $\rho_{0_{i}^{+}0_{1}^{+}}^{\mathrm{A}}$ is its nodal structure. If one disregards the nodes below about 1.5~fm for $i=2$ and 5, which stem from more complicated many-body properties, $\rho_{0_{i}^{+}0_{1}^{+}}^{\mathrm{A}}$ is characterized by one node at around the nuclear radius of $^{24}$Mg. As it is well known, this nodal structure is robust because of the orthogonal property:
\begin{equation}
\int \rho_{0_{i}^{+}0_{1}^{+}}^{\mathrm{A}}(r_{\mathrm{A}})
r_{\mathrm{A}}^2 dr_{\mathrm{A}}=0,
\quad (i\neq 1).
\end{equation}

With PWBA-$t$, the cross section is essentially determined by Eq.~(\ref{rhoatil}). If $q$ is very small, the difference between $j_0$ and unity is small and $j_0$ does not change its sign in the relevant region for Eq.~(\ref{rhoatil}). Thus, the contribution of $\rho_{0_{i}^{+}0_{1}^{+}}^{\mathrm{A}}$ at $r$ smaller than the node at $r_{\rm N}$ cancels with that for $r>r_{\rm N}$. To have a large cross section, the cancellation between $r<r_{\rm N}$ and $r>r_{\rm N}$ should be minimized. This is how the peak position (angle) of the cross section is determined in Figs.~\ref{fig3}(a) and \ref{fig4}(a). When a $g$ matrix is used instead of a $t$ matrix, an extra $r_{\rm A}$ dependence appears from the density dependence of the $g$ matrix, that is, the reduction of the NN effective interaction at finite densities. This makes the cancellation small and the cross section becomes large at very forward angles. A further hindrance of the cancellation is realized when the distortion effect, the nuclear absorption in particular, is included in the DWBA-$g$ calculation. Note that in the latter two cases, the leading term of $j_0$, which was dropped in the discussion in \S\ref{sec21}, can contribute to the cross section, which is responsible for making the cross section at $\theta=0^\circ$ maximum.

An important conclusion is that the $\alpha$ inelastic process is affected significantly by the nuclear distortion as well as the in-medium modification to the NN effective interaction. In other words, the explanation of the $B({\rm IS0})$-$(\alpha,\alpha')$ correspondence in the PW-LW limit does not make sense. It is shown that the $\alpha$ inelastic scattering process is not a process expressed by a single IS0 transition operator.

\subsection{Correspondence between the IS0 transition strength and $\alpha$ inelastic cross section}
\label{sec33}

%
\begin{table}[hbtp]
 \caption{IS0 transition strengths and the $\alpha$ inelastic cross sections relative to those for the $0^+_2$ state.}
 \label{tab2}
 \centering
 \begin{tabular}{cc|cccc}
  Quantity & reaction model & $0_3^+$ & $0_5^+$ & $0_7^+$ & $0_8^+$ \\
  \hline
  $B({\rm IS0})_i/B({\rm IS0})_2$                              & ------   & 0.33 & 0.18 & 0.09 & 0.42 \\
  \hline
                                                               & ${\rm PWBA}$-$t$ & 0.34 & 0.18 & 0.12 & 0.54 \\
  $({d\sigma_i}/{d\Omega})/({d\sigma_2}/{d\Omega})$ at 130~MeV & ${\rm PWBA}$-$g$ & 0.32 & 0.17 & 0.09 & 0.46 \\
                                                               & ${\rm DWBA}$-$g$ & 0.26 & 0.13 & 0.03 & 0.15 \\
  \hline
                                                               & ${\rm PWBA}$-$t$ & 0.34 & 0.18 & 0.12 & 0.53 \\
  $({d\sigma_i}/{d\Omega})/({d\sigma_2}/{d\Omega})$ at 386~MeV & ${\rm PWBA}$-$g$ & 0.33 & 0.17 & 0.09 & 0.46 \\
                                                               & ${\rm DWBA}$-$g$ & 0.30 & 0.15 & 0.06 & 0.29 \\
 \end{tabular}
\end{table}
%
Despite the apparently negative conclusion on the $B({\rm IS0})$-$(\alpha,\alpha')$ correspondence drawn in \S\ref{sec32}, there have been many studies in which $B({\rm IS0})$ were {\it successfully} extracted from $\alpha$ inelastic scattering data. To see the situation in the present case, we show in Table~\ref{tab2} the relative IS0 strength to that for the $0_2^+$ state, $B({\rm IS0})_i/B({\rm IS0})_2$, calculated with AMD and the relative $\alpha$ inelastic cross sections, $({d\sigma_i}/{d\Omega})/({d\sigma_2}/{d\Omega})$, at 130 and 386~MeV, evaluated with PWBA-$t$, PWBA-$g$, and DWBA-$g$. The cross sections at the peaks around $8^\circ$ and $5^\circ$ are used at 130 and 386~MeV, respectively. For DWBA-$g$, we have evaluated the ratios also at $0^\circ$, which are found to be almost identical to the values in Table~\ref{tab2} (not shown).

One can see that the behavior of $({d\sigma_i}/{d\Omega})/({d\sigma_2}/{d\Omega})$ for the $0_3^+$ and $0_5^+$ states is somewhat different from that for the $0_7^+$ and $0_8^+$ states. For the former, the PWBA-$t$ and PWBA-$g$ results show very good agreement with $B({\rm IS0})_i/B({\rm IS0})_2$ at both energies. Although the DWBA-$g$ results slightly deviate from the $B({\rm IS0})$ ratios, the $B({\rm IS0})$-$(\alpha,\alpha')$ correspondence holds with an error of less than about $20\%$--$30\%$ ($10\%$--$15\%$) at 130 (386)~MeV. This suggests a {\lq\lq}robustness'' of the $B({\rm IS0})$-$(\alpha,\alpha')$ correspondence against the strong distortion and in-medium effects shown in \S\ref{sec32}. On the other hand, for the latter, the cross section ratios for PWBA-$t$ are different from the $B({\rm IS0})$ ratios by about $25\%$. Furthermore, the difference between the PWBA-$t$, PWBA-$g$, and DWBA-$g$ results is significantly larger than that for the $0_3^+$ and $0_5^+$ states. For the the $0_7^+$ and $0_8^+$ states, we conclude that the $B({\rm IS0})$-$(\alpha,\alpha')$ correspondence does not hold well. Below we investigate the mechanism of the the $B({\rm IS0})$-$(\alpha,\alpha')$ correspondence for the $0_3^+$ and $0_5^+$ states and the breakdown of it for the $0_7^+$ and $0_8^+$ states one by one.

As discussed in \S\ref{sec32}, the transition densities between $0^+$ states are constrained rather strongly, and can be well described by a macroscopic model~\cite{Sat87}:
\begin{equation}
\rho_{0_{i}^{+}0_{1}^{+}}^{\mathrm{A}}=-\alpha_{0i} \left(3+r_{\rm A} \frac{d}{dr_{\rm A}}\right) \rho_0(r_{\rm A}),
\label{tdm}
\end{equation}
where $\rho_{0}$ is the diagonal density for the ground state. $\alpha_{0i}$ is a dimensionless deformation parameter that can be evaluated by an energy-weighted sum rule if one excited state exhausts the total transition strength. In the following discussion, however, $\alpha_{0i}$ is regarded as just a normalization parameter that characterizes the IS0 transition strength to the $i$th state and is expected to be determined by the analysis of $(\alpha,\alpha')$ cross section data.

Once the functional form of Eq.~(\ref{tdm}) is assumed, trivially, we have
\begin{equation}
\left(
\displaystyle\frac
{\int \rho_{0_i^+ 0_1^+}^{\rm A} F(r_{\rm A}) d{\bm r}_{\rm A}}
{\int \rho_{0_j^+ 0_1^+}^{\rm A} F(r_{\rm A}) d{\bm r}_{\rm A}}
\right)^2
=
\left(
\displaystyle\frac{\alpha_{0i}}{\alpha_{0j}}
\right)^2
\label{macro}
\end{equation}
for any function $F$ of $r_{\rm A}$ that has no $i$ dependence. If $F$ is $r_{\rm A}^2$, the left-hand-side of Eq.~(\ref{macro}) means the $B({\rm IS0})$ ratio, whereas it becomes the ratio of the cross sections obtained with PWBA-$t$ if $F(r_{\rm A})=j_0 (q r_{\rm A})$. Even when an extra $r_{\rm A}$ dependence coming from the $g$ matrix or nuclear absorption appears, Eq.~(\ref{macro}) holds when the $r_{\rm A}$ dependence is the same for all the states. This is the reason why $B({\rm IS0})_i/B({\rm IS0})_2$ tends to keep a clear correspondence with $({d\sigma_i}/{d\Omega})/({d\sigma_2}/{d\Omega})$ even though the simple explanation in \S\ref{sec21} does not hold in the actual $\alpha$ inelastic scattering. In short, when the transition densities are given by Eq.~(\ref{tdm}) and the distortion and in-medium effects have no channel dependence, the $B({\rm IS0})$-$(\alpha,\alpha')$ correspondence holds as long as the CC effect is disregarded.

%
\begin{figure}[htpb]
\begin{center}
\includegraphics[width=0.6\textwidth,clip]{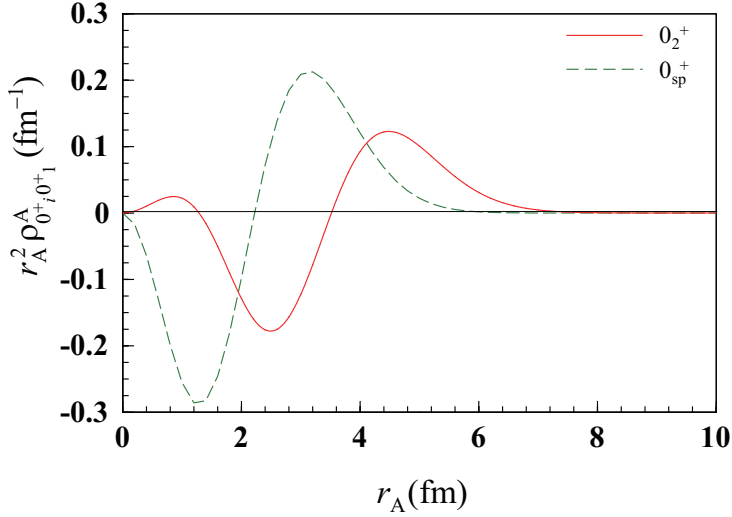}
\caption{Transition densities multiplied by $r_{\rm A}^2$ calculated with the s.p. H.O. model (dashed). The solid line is the same as in Fig.~\ref{fig5}.}
\label{fig6}
\end{center}
\end{figure}
Next we discuss the breakdown of the $B({\rm IS0})$-$(\alpha,\alpha')$ correspondence for the $0_7^+$ and $0_8^+$ states. As seen from Fig.~\ref{fig5}, the positions of the node, $r_{\rm N}$, of the transition densities for $i=7$ and 8 are smaller than for the other states. Obviously, this difference stains the $B({\rm IS0})$-$(\alpha,\alpha')$ correspondence. It should be noted that the difference in the transition density causes different $r_{\rm A}$ dependence of $F$ in Eq.~(\ref{macro}) due to the distortion and density-dependent $g$ matrix. To confirm this, a transition density based on a single-particle (s.p.) model is prepared. We assume an s.p. transition from $0s$ to $1s$ orbitals in a harmonic oscillator (H.O.) potential; the H.O. parameter $\hbar \omega$ is taken to be 12.7~MeV, which reproduces the root-mean-square radius of the ground state of $^{24}$Mg calculated with AMD. Henceforth, we denote this transition density $\rho_{0_{\rm sp}^{+}0_{1}^{+}}^{\mathrm{A}}$. In Fig.~\ref{fig6} we show $r_{\rm A}^2 \rho_{0_{\rm sp}^{+}0_{1}^{+}}^{\mathrm{A}}$ by the dashed line; $r_{\rm A}^2 \rho_{0_{2}^{+}0_{1}^{+}}^{\mathrm{A}}$ is also shown by the solid line for comparison. We have renormalized $\rho_{0_{\rm sp}^{+}0_{1}^{+}}^{\mathrm{A}}$ so that the $B({\rm IS0})$ ratio to the $0_2^+$ state is unity. One sees a significant difference in $r_{\rm N}$ between the two densities. Reflecting this, the cross section ratios at 386~MeV are found to be 2.18 (PWBA-$t$), 1.59 (PWBA-$g$), and 0.52 (DWBA-$g$). Therefore, for excited states having a different structural nature, the $B({\rm IS0})$-$(\alpha,\alpha')$ correspondence will not hold even at 386~MeV.

From a detailed analysis of the AMD wave function, it is found that the $0_7^+$ and $0_8^+$ states contain significant components of the ${}^{12}{\rm C}+{}^{12}{\rm C}$ configuration, whereas the $\alpha+{}^{20}$Ne configuration is dominant for the other $0^+$ states. The excitation of $^{24}$Mg to a state having the $\alpha+{}^{20}$Ne configuration can be interpreted as a four-nucleon excitation from the $sd$-orbits in terminology of the na\"ive shell model. On the other hand, the excitation to the $0_7^+$ and $0_8^+$ states corresponds to excitation of nucleons from the $p$-orbits. Under the condition that the transition density must have a node, therefore, $r_{\rm N}$ becomes small for a transition to a state having a ${}^{12}{\rm C}+{}^{12}{\rm C}$ configuration. It will be interesting that $r_{\rm N}$ can be an indicator of an intrinsic structure of the $0^+$ excited states. A structure model that does not assume a specific cluster structure {\it a priori} combined with a microscopic reaction framework will be very important to discuss the development of cluster states and its correspondence with reaction observables.

\subsection{Coupled-channel effect on the $\alpha$ inelastic cross section}
\label{sec34}

%
\begin{figure}[htpb]
\begin{center}
\includegraphics[width=0.48\textwidth,clip]{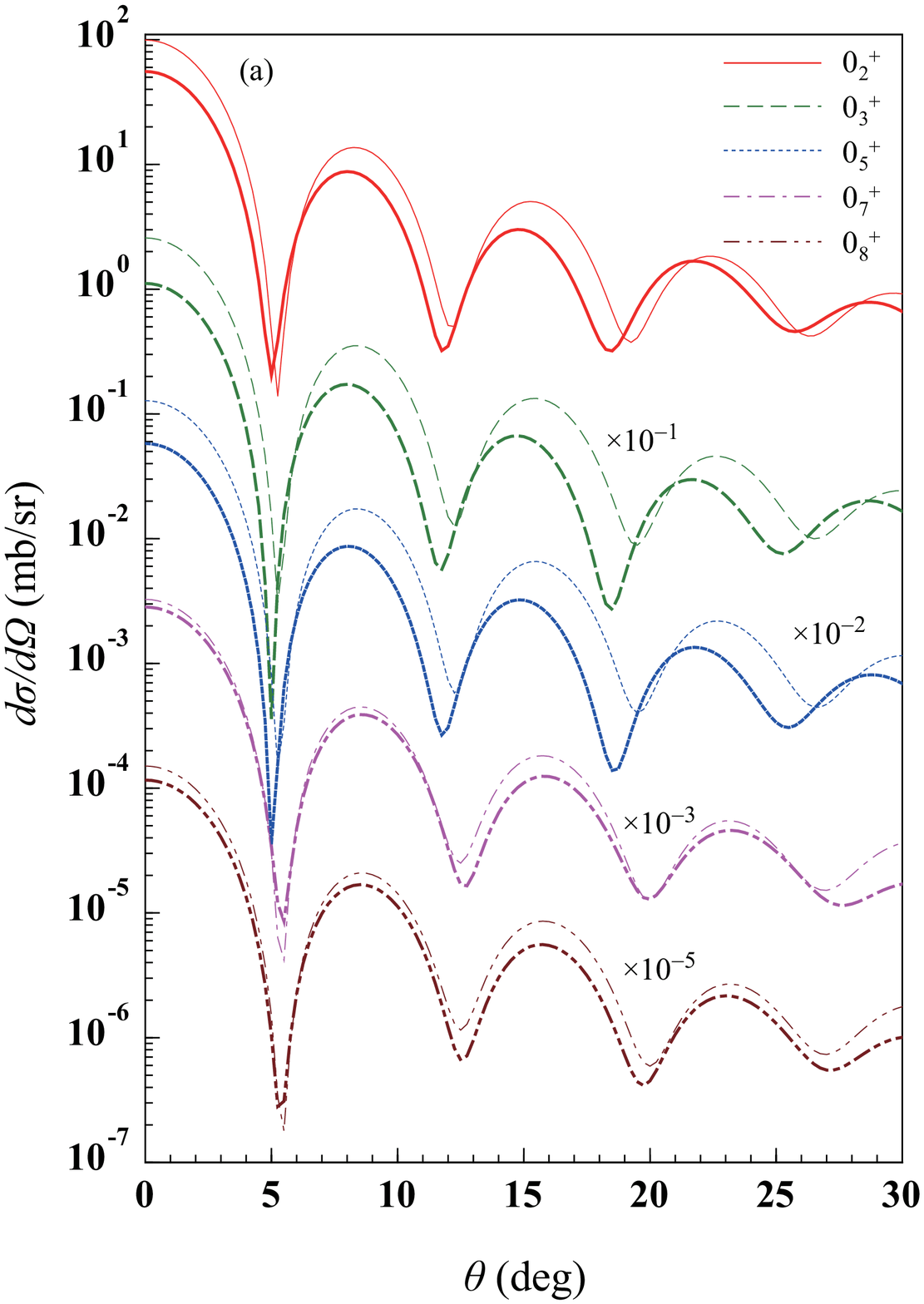}
\includegraphics[width=0.48\textwidth,clip]{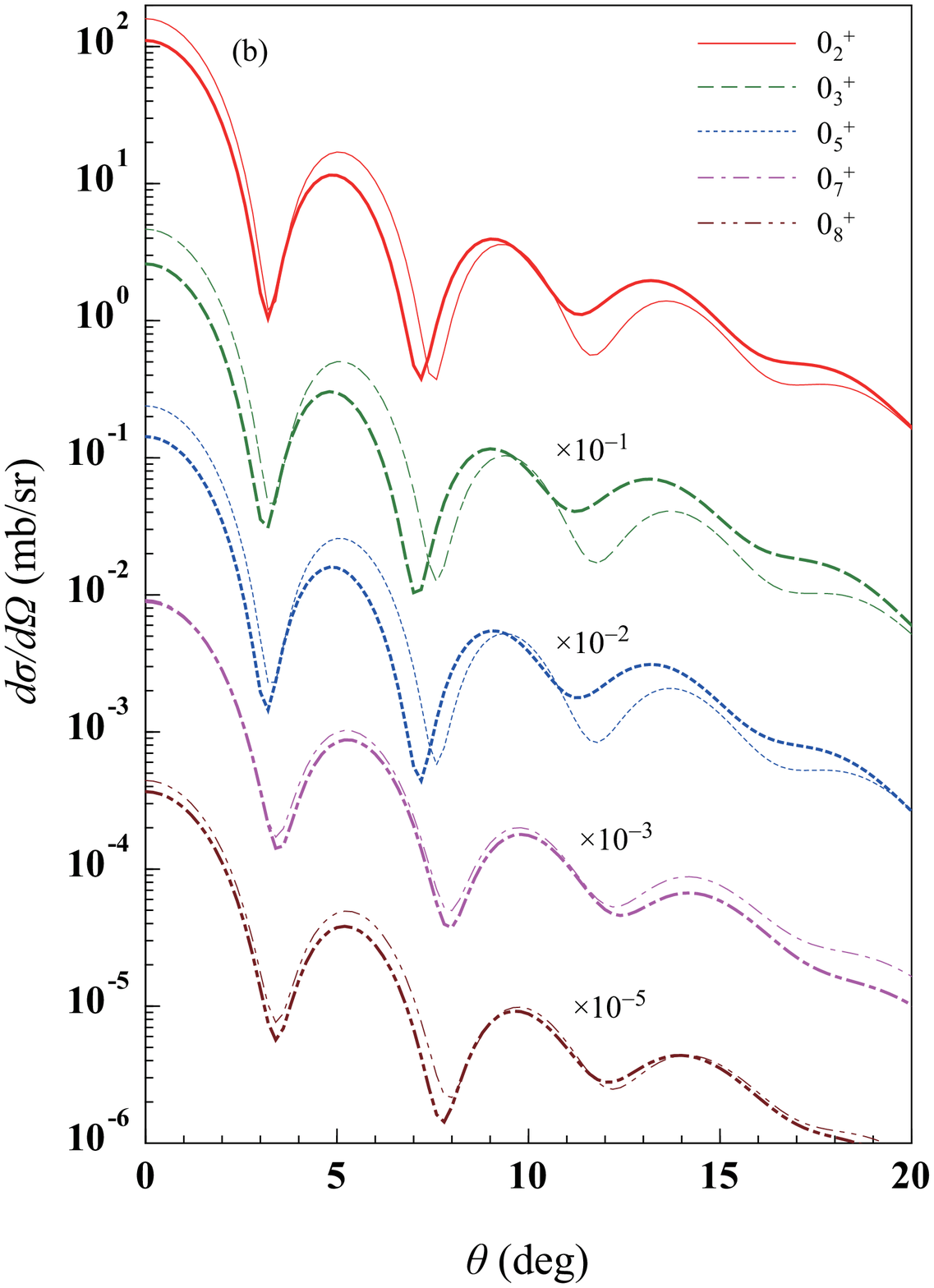}
\caption{$\alpha$ inelastic cross section on $^{24}$Mg at (a) 130 and (b) 386~MeV obtained with the CC (thick lines) and DWBA (thin lines) calculations.}
\label{fig7}
\end{center}
\end{figure}
As the last piece that potentially stains the $B({\rm IS0})$-$(\alpha,\alpha')$ correspondence, we discuss the CC effect in the $\alpha$ inelastic scattering.
In Fig.~\ref{fig7}, we show the $\alpha$ inelastic cross sections at (a) 130 and  (b) 386~MeV as a function of $\theta$. The meaning of the lines is the same as in Fig.~\ref{fig3} and the thick (thin) lines show the results of the CC (DWBA) calculation. The Melbourne $g$ matrix is employed in all the cases. One sees that, as indicated in Figs.~\ref{fig1}(b) and \ref{fig2}(b), a CC effect on the cross sections remains non-negligible at both energies.

%
\begin{table}[hbtp]
 \caption{Ratio of the $\alpha$ inelastic cross sections calculated with the CC calculation to those with DWBA.}
 \label{tab3}
 \centering
 \begin{tabular}{cc|ccccc}
   energy & angle             & $0_2^+$ & $0_3^+$ & $0_5^+$ & $0_7^+$ & $0_8^+$ \\
  \hline
  130~MeV & first peak ($0^\circ$)       & 0.56    & 0.43    & 0.45    & 0.87    & 0.77    \\
          & second peak ($\sim 5^\circ$) & 0.64    & 0.49    & 0.50    & 0.87    & 0.80    \\
  386~MeV & first peak ($0^\circ$)       & 0.69    & 0.56    & 0.60    & 0.97    & 0.83    \\
          & second peak ($\sim 8^\circ$) & 0.68    & 0.60    & 0.62    & 0.85    & 0.78    \\
 \end{tabular}
\end{table}
%
The ratios of ${d\sigma_i}/{d\Omega}$ obtained with the CC calculation to those with DWBA, determined at $\theta=0^\circ$ and at the second peak, are listed in Table~\ref{tab3}. One sees the deviation from unity is around 30\%--60\% for $i=$2, 3, and 5. On the other hand, the deviation is rather small for $i=7$ and 8. The small CC effect on the cross sections for $i=7$ and 8 may be due to the ${}^{12}{\rm C}+{}^{12}{\rm C}$ configuration for these states. It will be worth pointing out that the energy dependence of the CC effect is not so strong when the results at 130~MeV and 386~MeV are compared. It is found by a detailed investigation that the coupling through the $2_1^+$ state is dominant for the CC effect on the inelastic cross sections to the $0^+$ states; for the $0_2^+$ cross sections, the coupling through the $2_2^+$ state also plays a role. It should be noted, however, that in the present study we included only three low-lying $2^+$ states. A more complete CC calculation including other $2^+$ states as well as $4^+$, $1^-$, and $3^-$ states will be needed to draw a definite conclusion on the CC effect.

%
\begin{table}[hbtp]
 \caption{Same as Table~\ref{tab2} but for the CC calculation evaluated at $0^\circ$ and the second peak. The results of $B({\rm IS0})_i/B({\rm IS0})_2$ and those corresponding to DWBA-$g$ are also shown for comparison.}
 \label{tab4}
 \centering
 \begin{tabular}{cc|cccc}
  Quantity & reaction model & $0_3^+$ & $0_5^+$ & $0_7^+$ & $0_8^+$ \\
  \hline
  $B({\rm IS0})_i/B({\rm IS0})_2$                              & ------                   & 0.33 & 0.18 & 0.09 & 0.42 \\
  \hline
                                                               & ${\rm DWBA}$-$g$         & 0.26 & 0.13 & 0.03 & 0.15 \\
  $({d\sigma_i}/{d\Omega})/({d\sigma_2}/{d\Omega})$ at 130~MeV & ${\rm CC}$ ($0^\circ$)   & 0.20 & 0.10 & 0.05 & 0.21 \\
                                                               & ${\rm CC}$ (second peak) & 0.20 & 0.10 & 0.04 & 0.19 \\
  \hline
                                                               & ${\rm DWBA}$-$g$         & 0.30 & 0.15 & 0.06 & 0.29 \\
  $({d\sigma_i}/{d\Omega})/({d\sigma_2}/{d\Omega})$ at 386~MeV & ${\rm CC}$ ($0^\circ$)   & 0.23 & 0.13 & 0.08 & 0.33 \\
                                                               & ${\rm CC}$ (second peak) & 0.26 & 0.14 & 0.08 & 0.33 \\
 \end{tabular}
\end{table}
%
The state dependence of the values in Table~\ref{tab3} affects the $B({\rm IS0})$-$(\alpha,\alpha')$ correspondence. In Table~\ref{tab4}, we show the ratio of the cross sections obtained with the CC calculation to that to the $0_2^+$ state; $B({\rm IS0})_i/B({\rm IS0})_2$ and the results for DWBA-$g$, which are the same as in Table~\ref{tab2}, are also shown for comparison. One sees that the $B({\rm IS0})$-$(\alpha,\alpha')$ correspondence at 386~MeV becomes worse for $i=3$ but is still within an error of about 20\%. On the other hand, an improvement of the result is found for $i=7$ and 8, though it should be considered to be accidental. At 130~MeV, the cross section ratio deviates from the $B({\rm IS0})$ ratio by $40\%$--$55\%$. It will be, therefore, difficult to reliably extract $B({\rm IS0})$ from $(\alpha,\alpha')$ cross section data at 130~MeV.

\section{Summary}
\label{sum}

We have investigated the correspondence between the isoscalar monopole (IS0) transition strengths and $\alpha$ inelastic cross sections, the $B({\rm IS0})$-$(\alpha,\alpha')$ correspondence, for $^{24}$Mg($\alpha,\alpha'$) at 130 and 386~MeV. We prepared diagonal and transition densities of $^{24}$Mg with AMD and performed a microscopic coupled-channel calculation based on a folding model for reaction observables. The calculated elastic cross section and inelastic cross sections to the $0^+_2$ and $2_1^+$ states are found to reasonably reproduce the experimental data with no adjustable parameter in the reaction calculation.

We found that the $\alpha$ inelastic cross sections are significantly affected by the nuclear distortion and the in-medium modification to the nucleon-nucleon effective interaction, and cannot be regarded as a process expressed by a single IS0 transition operator. Nevertheless, the $B({\rm IS0})$-$(\alpha,\alpha')$ correspondence tends to remain, at least to some extent, because of a rather strong constraint on the monopole transition densities. However, when a $0^+$ state characterized by a different intrinsic structure is considered, the $B({\rm IS0})$-$(\alpha,\alpha')$ correspondence is stained. Besides, the coupled-channel effect is found to be non-negligible at not only 130~MeV but also 386~MeV. Mainly because of the strong nuclear distortion, the $B({\rm IS0})$-$(\alpha,\alpha')$ correspondence does not hold at 130~MeV. These findings will be expected to be also the case with the $(\alpha,\alpha')$ scattering to $0^+$ states of other nuclei. Characteristics of individual $0^+$ states, whether a macroscopic model is valid or not, of course depend on nuclei.

In conclusion, the $\alpha$ inelastic scattering is shown to be a process that cannot be expressed by a single IS0 transition operator. Even though the $B({\rm IS0})$-$(\alpha,\alpha')$ correspondence holds reasonably well after all at 386~MeV, with an error of about 20\%--30\%, it will be misleading to explain the correspondence in the plane wave limit combined with the long-wavelength approximation. When a $0^+$ excited state that has an exotic configuration compared to the standard $\alpha$ cluster state is considered, the $B({\rm IS0})$-$(\alpha,\alpha')$ correspondence tends to be questionable. Such states cannot be differentiated from others when a macroscopic model is adopted. A microscopic description of both the structure and reaction parts will be very important to discuss the variety of cluster states of nuclei with reaction observables. Systematic investigation on other nuclei and on inelastic processes to other spin-parity states, $1^-$ and $2^+$ states in particular, will be important future work.

\section*{Acknowledgments}

The authors thank Y.~Kanada-En'yo, M.~Kimura, Y.~Taniguchi, K.~Yoshida, and K.~Sato for fruitful discussions. This work was supported in part by Grants-in-Aid of the Japan Society for the Promotion of Science (Grants No. JP16K05352 and No. JP20829832) and by the COREnet program of RCNP, Osaka University.

\end{document}